\documentclass[preprint,aps,prl,floats,a4paper,superscriptaddress,
preprintnumbers,showpacs,tightenlines]{revtex4}
\draft 
\usepackage{amsmath}
\usepackage{graphicx}
\usepackage{epsfig}

\usepackage{dcolumn}
\newcolumntype{B}{D{B}{}{-1}}

\newcommand{\bzbar}{\ensuremath{\overline{B}{}^0}}
\newcommand{\bbar}{\ensuremath{\overline{B}}}
\newcommand{\dbar}{\ensuremath{\overline{D}}}
\newcommand{\kzbarstar}{\ensuremath{\overline{K}{}^{*0}}}

\newcommand{\mev}{\ensuremath{\mathrm{MeV}}}
\newcommand{\gev}{\ensuremath{\mathrm{GeV}}}

\begin{document}

\setlength{\unitlength}{1pt}

\preprint{\tighten\vbox{\hbox{\hfil Belle Prerpint 2004-30}
                        \hbox{\hfil KEK   Preprint 2004-59}
\vskip -2.0cm}}

\noindent
\epsfysize3cm
\epsfbox{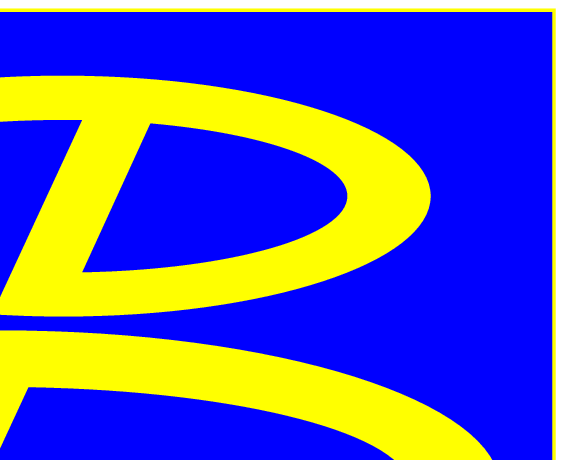}    

\vskip 0.5cm

\title{
\bf Observation of 
{\boldmath $\bzbar$} $\rightarrow$ {\boldmath $D$}$\,^\ast$\hspace{-0.2cm}{\boldmath $_{\it sJ}(2317)^+ K$}$^-$ decay.}

\affiliation{Budker Institute of Nuclear Physics, Novosibirsk}
\affiliation{Chiba University, Chiba}
\affiliation{Chonnam National University, Kwangju}
\affiliation{University of Cincinnati, Cincinnati, Ohio 45221}
\affiliation{University of Frankfurt, Frankfurt}
\affiliation{Gyeongsang National University, Chinju}
\affiliation{University of Hawaii, Honolulu, Hawaii 96822}
\affiliation{High Energy Accelerator Research Organization (KEK), Tsukuba}
\affiliation{Hiroshima Institute of Technology, Hiroshima}
\affiliation{Institute of High Energy Physics, Chinese Academy of Sciences, Beijing}
\affiliation{Institute of High Energy Physics, Vienna}
\affiliation{Institute for Theoretical and Experimental Physics, Moscow}
\affiliation{J. Stefan Institute, Ljubljana}
\affiliation{Korea University, Seoul}
\affiliation{Kyungpook National University, Taegu}
\affiliation{Swiss Federal Institute of Technology of Lausanne, EPFL, Lausanne}
\affiliation{University of Ljubljana, Ljubljana}
\affiliation{University of Maribor, Maribor}
\affiliation{University of Melbourne, Victoria}
\affiliation{Nagoya University, Nagoya}
\affiliation{Nara Women's University, Nara}
\affiliation{National Central University, Chung-li}
\affiliation{National United University, Miao Li}
\affiliation{Department of Physics, National Taiwan University, Taipei}
\affiliation{H. Niewodniczanski Institute of Nuclear Physics, Krakow}
\affiliation{Nihon Dental College, Niigata}
\affiliation{Niigata University, Niigata}
\affiliation{Osaka City University, Osaka}
\affiliation{Osaka University, Osaka}
\affiliation{Panjab University, Chandigarh}
\affiliation{Peking University, Beijing}
\affiliation{Princeton University, Princeton, New Jersey 08545}
\affiliation{Saga University, Saga}
\affiliation{University of Science and Technology of China, Hefei}
\affiliation{Seoul National University, Seoul}
\affiliation{Sungkyunkwan University, Suwon}
\affiliation{University of Sydney, Sydney NSW}
\affiliation{Tata Institute of Fundamental Research, Bombay}
\affiliation{Toho University, Funabashi}
\affiliation{Tohoku Gakuin University, Tagajo}
\affiliation{Tohoku University, Sendai}
\affiliation{Department of Physics, University of Tokyo, Tokyo}
\affiliation{Tokyo Institute of Technology, Tokyo}
\affiliation{Tokyo Metropolitan University, Tokyo}
\affiliation{Tokyo University of Agriculture and Technology, Tokyo}
\affiliation{University of Tsukuba, Tsukuba}
\affiliation{Virginia Polytechnic Institute and State University, Blacksburg, Virginia 24061}
\affiliation{Yonsei University, Seoul}
  \author{A.~Drutskoy}\affiliation{University of Cincinnati, Cincinnati, Ohio 45221} 
  \author{K.~Abe}\affiliation{High Energy Accelerator Research Organization (KEK), Tsukuba} 
  \author{K.~Abe}\affiliation{Tohoku Gakuin University, Tagajo} 
  \author{I.~Adachi}\affiliation{High Energy Accelerator Research Organization (KEK), Tsukuba} 
  \author{H.~Aihara}\affiliation{Department of Physics, University of Tokyo, Tokyo} 
  \author{M.~Akatsu}\affiliation{Nagoya University, Nagoya} 
  \author{Y.~Asano}\affiliation{University of Tsukuba, Tsukuba} 
  \author{T.~Aushev}\affiliation{Institute for Theoretical and Experimental Physics, Moscow} 
  \author{S.~Bahinipati}\affiliation{University of Cincinnati, Cincinnati, Ohio 45221} 
  \author{A.~M.~Bakich}\affiliation{University of Sydney, Sydney NSW} 
% ...
\collaboration{The Belle Collaboration}

\date{\today}

\tighten

\begin{abstract}
The decays \mbox{$\bzbar \rightarrow D_{sJ}^+ K^-$} and 
\mbox{$\bzbar \rightarrow D_{sJ}^- \pi^+$} are studied for
the first time.
A significant signal is observed in the
$\bzbar \rightarrow D_{sJ}^\ast(2317)^+ K^-$ decay channel with
${\cal B}(\bzbar \rightarrow D_{sJ}^\ast(2317)^+ K^-) \times
{\cal B} (D_{sJ}^\ast(2317)^+ \rightarrow D_s^+ \pi^0) =
(5.3^{+1.5}_{-1.3} \pm 0.7 \pm 1.4) \times 10^{-5}$.
No signals are observed in the $\bzbar \rightarrow D_{sJ}^\ast(2317)^- \pi^+$,
$\bzbar \rightarrow D_{sJ}(2460)^+ K^-$, and 
$\bzbar \rightarrow D_{sJ}(2460)^- \pi^+$ decay modes,
and upper limits are obtained.
The analysis is based on a data set of 140~fb$^{-1}$ collected 
by the Belle experiment at the asymmetric $e^+ e^-$ collider KEKB.
\end{abstract}


\maketitle

Two narrow resonances denoted as $D_{sJ}^\ast(2317)^+$ and $D_{sJ}(2460)^+$
have been observed recently 
in $e^+ e^-$ continuum interactions~\cite{baba,cleo,bela,babb}.
These resonances were initially seen in the $D_{sJ}^\ast(2317)^+ \to D_s^+ \pi^0
$,
$D_{sJ}(2460)^+ \to D_s^+ \gamma$ and $D_{sJ}(2460)^+\to D_s^{*+} \pi^0$
decay modes \cite{fna}, and
their quantum numbers were tentatively classified
as $J^P=0^+$ for $D_{sJ}^\ast(2317)^+$ and $J^P=1^+$ for $D_{sJ}(2460)^+$.
However, the measured masses are significantly lower
than the values predicted within potential models for $0^+$ 
and $1^+$ states~\cite{dsjaa}.
The $D_{sJ}$ mesons were also observed
in $B \to \dbar D_{sJ}$ decay modes with branching fractions
an order of magnitude less than those for 
$B \to \dbar D_{s}$ decay modes with a pseudoscalar $D_s$~\cite{belb}.
Angular analysis of $B \to \dbar D_{sJ}(2460)^+$ 
favors a spin 1 assignment for $D_{sJ}(2460)^+$.
There has been a significant effort to explain the surpising
$D_{sJ}$ masses~\cite{dsjaa},
and some authors have discussed the possibility of
four-quark content 
in the $D_{sJ}^+$~\cite{dsje,dsjf,dsjg,dsjh,dsji}.

In this Letter we report the results from a search for
$\bzbar \to D_{sJ}^+ K^-$ and $\bzbar \to D_{sJ}^- \pi^+$ decays,
where $D_{sJ}^+$ mesons
are reconstructed in the modes $D_{sJ}^\ast(2317)^+ \to D_s^+ \pi^0$
and $D_{sJ}(2460)^+ \to D_s^+ \gamma$.
Measurements of 
the corresponding decays $\bzbar \to D_s^+ K^-$ and 
$\bzbar \to D_s^- \pi^+$ have been reported recently
by Belle~\cite{belc} and BaBar~\cite{babc}.

The decay mode $\bzbar \to D_{s(J)}^- \pi^+$ can be described by
a $b\to u$ tree diagram. Within the factorization
approach~\cite{pteoa}, the branching fraction ratio
$R_{\pi^+/D^+} = {\cal B}(\bzbar \to D_s^- \pi^+) / {\cal B}(\bzbar \to D_s^- D^+)$
is predicted to be $(0.424 \pm 0.041) \cdot |V_{ub}/V_{cb}|^2$ and
can be used to obtain the ratio of Cabbibo-Kobayashi-Maskawa 
matrix elements $|V_{ub}/V_{cb}|$. Assuming
similar ratios $R_{\pi^+/D^+}$ for
$D_s$ and $D_{sJ}$ mesons,
only a few $\bzbar \to D_{sJ}^- \pi^+$ events would be
observed in the current Belle data sample.

\begin{figure}[h!]
\vspace{-0.9cm}
\begin{center}
\epsfig{file=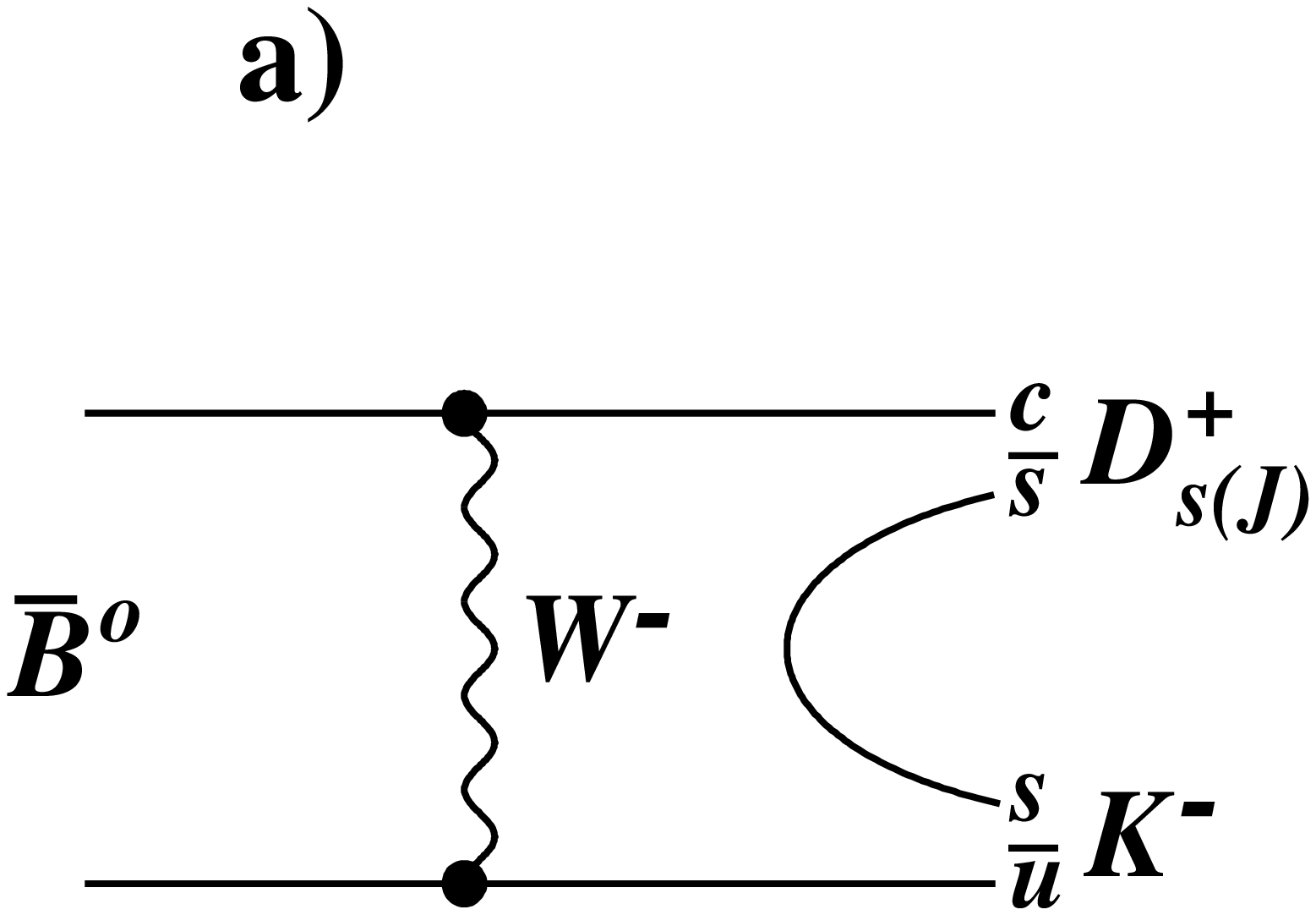,width=5.cm,height=5.cm}\epsfig{file=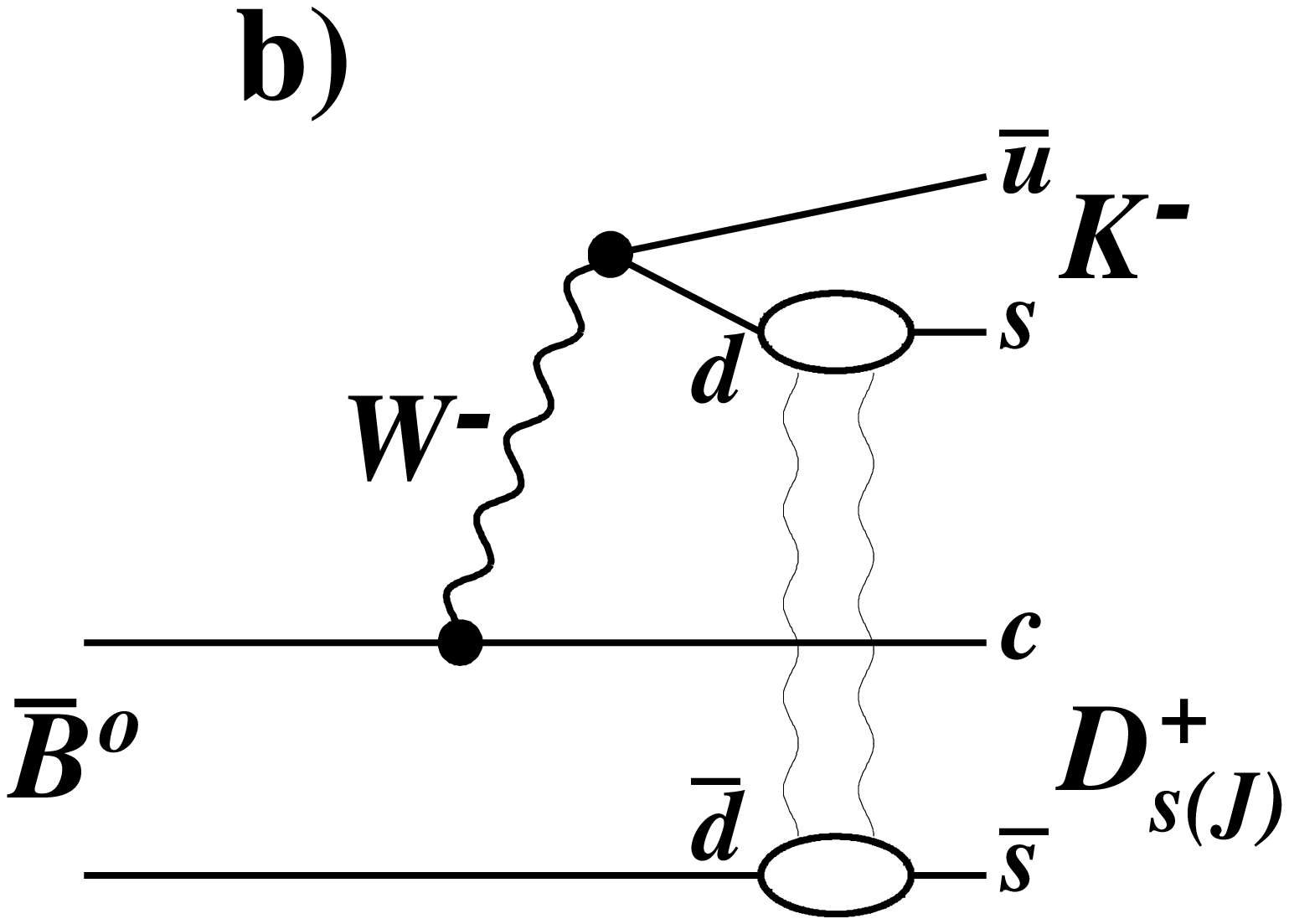,width=5.cm,height=5.cm}\epsfig{file=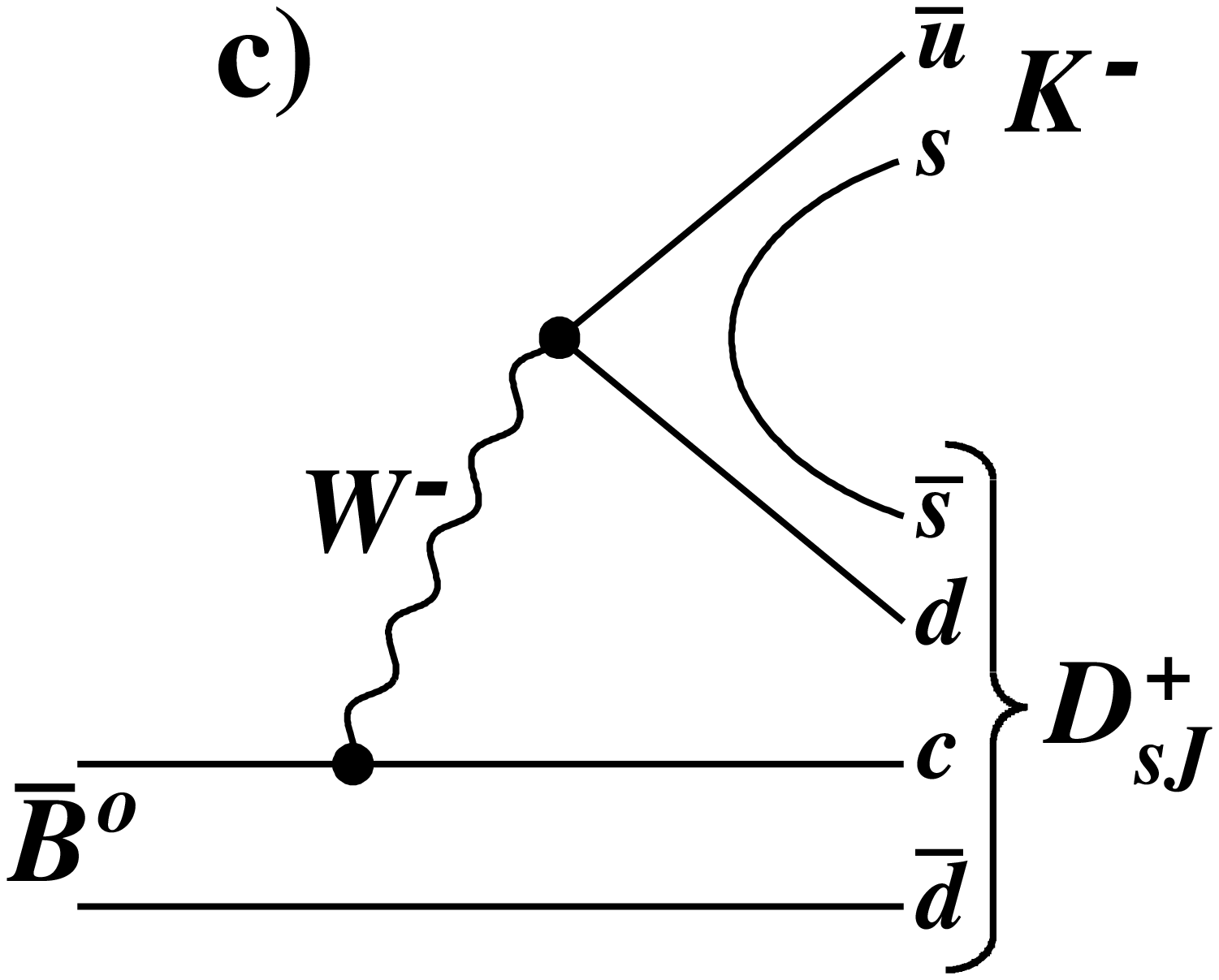,width=5.cm,height=5.cm}
\end{center}
\vspace{-0.7cm}
\caption{ Diagrams describing $\bzbar \to D_{sJ}^+ K^-$ decay.}
\label{diagramms}
\end{figure}

The decays $\bzbar \to D_{s(J)}^+ K^-$ are of special interest because 
the quark content of the initial \bzbar\ meson ($b\bar{d}$)
is completely different from that of the $D_{s(J)}^+ K^-$ 
final state ($cs\bar{s}\bar{u}$),
indicating an unusual configuration with both initial quarks 
involved in the weak decay.
Branching fractions for the pseudoscalar $D_s^+$ meson
${\cal B}(\bzbar \to D_s^+ K^-) = (4.6^{+1.2}_{-1.1} \pm 1.3)\cdot 10^{-5}$ and
\mbox{$(3.2 \pm 1.0 \pm 1.0)\cdot 10^{-5}$}
were measured by the Belle~\cite{belc} and BaBar~\cite{babc} collaborations, respectively.
Predictions for this branching fraction have been obtained
assuming a dominant contribution from a PQCD factorization $W$ 
exchange process (Fig.~\ref{diagramms}a)~\cite{kteoa,kteob}
or, alternatively, from final state interactions (Fig.~\ref{diagramms}b)~\cite{kteoc,kteod},
and cover the range from a few units times $10^{-6}$ to $10^{-4}$. 
If the $D_{sJ}$ mesons have a four-quark component
then the tree diagram with $s\bar{s}$ pair 
creation (shown in Fig.~\ref{diagramms}c) may also contribute.

The analysis was performed using a $140\,\mathrm{fb}^{-1}$ data sample
containing $(152.0 \pm 0.7)\times 10^6$ $B\bbar$ pairs. The data were
collected with the Belle detector at KEKB \cite{kekb},
an asymmetric energy double storage ring collider with 8 GeV electrons 
and 3.5 GeV positrons.
Belle is a general-purpose large-solid-angle detector 
that consists of a three-layer Silicon Vertex Detector (SVD),
a 50-layer Central Drift Chamber (CDC), an array of 
Aerogel \v{C}erenkov Counters (ACC),  a Time of Flight
counter system (TOF), and a CsI(Tl) Electromagnetic 
Calorimeter (ECL) located inside a superconducting
solenoidal coil with a 1.5~T magnetic field. 
The detector is described in detail elsewhere~\cite{BELLE_DETECTOR}.

Charged tracks are required to have momentum $p > 100\,\mev/c$~\cite{fnb}
and impact parameters less than 2~cm radially
and 5~cm in the $z$ direction~\cite{zdef} with respect to the interaction point.
Kaon and pion mass hypotheses are assigned using a likelihood 
ratio ${\cal L}_{K/\pi} = {\cal L}_K/({\cal L}_K + {\cal L}_{\pi})$, obtained
by combining information from the CDC ($dE/dx$), ACC, and TOF systems.
We require ${\cal L}_{K/\pi} > 0.6$ (${\cal L}_{K/\pi} < 0.6$) 
for kaon (pion) candidates~\cite{BELLE_DETECTOR}. With these requirements
the identification efficiency for particles used in this analysis varies
from 91\% to 86\% for kaons and from 98\% to 94\% for pions,
decreasing as the momentum increases.

ECL clusters with a photonlike shape and energy larger than 50 MeV,
that are not associated with charged tracks,
are accepted as photon candidates. Photon pairs of invariant mass within
$\pm 12\,\mev/c^2$ (\mbox{$\sim\,3\sigma$} in the $\pi^0$ mass resolution) 
of the $\pi^0$ mass are considered $\pi^0$ candidates;
the $\pi^0$ momentum is required to be larger than $100\,\mev/c$.

$K^0_S$ candidates are formed from $\pi^+\pi^-$ pairs 
with an invariant mass within $\pm 10\,\mev/c^2$ ($\sim$\,3$\sigma$)
of the nominal $K^0_S$ mass and a common vertex
displaced from the interaction point by more than 0.2\,cm in the
plane perpendicular to the beam direction.
A common vertex for the two tracks in the plane perpendicular to the beam
direction was found; the difference in $z$ coordinates of the measured pion
tracks at this point was required to be less than 2\,cm.
The angle $\alpha$ between the $K^0_S$ flight and momentum directions
is required to satisfy $\cos\alpha\,>\,0.8$.

Invariant masses of $K^{*0} \to K^+ \pi^-$ candidates are required to
be within $\pm 50\,\mev/c^2$ of the nominal $K^{*0}$ mass; those of
$\phi \to K^+ K^-$ candidates, 
within $\pm 12\,\mev/c^2$ of the $\phi$ mass.
$D_s^+$ mesons are reconstructed in the $\phi \pi^+$, $\kzbarstar K^+$
and $K_S^0 K^+$ decay channels; a mass window of 
$\pm 12\,\mev/c^2$ ($\sim\,2.5\sigma$) is imposed in each case.

The $D_{sJ}$ mesons
are reconstructed in the $D_{sJ}^\ast(2317)^+ \to D_s^+ \pi^0$ and
$D_{sJ}(2460)^+ \to D_s^+ \gamma$ decay modes.
To select a $D_{sJ}^\ast(2317)^+$, the candidate mass difference 
$\Delta M(D_{sJ}^\ast(2317)^+) \equiv M(D_s^+ \pi^0) - M(D_s^+)$
is required to lie
within $\pm 20\,\mev/c^2$ of $348.6\,\mev/c^2$ ($\sim\,3.0\sigma$).
To select a $D_{sJ}(2460)^+$, we require
$\Delta M(D_{sJ}(2460)^+) \equiv M(D_s^+ \gamma) - M(D_s^+)$ within 
$\pm 30\,\mev/c^2$ of $487.9\,\mev/c^2$ ($\sim\,2.5\sigma$).
The $D_{sJ}$ and $D_{s}$ mass differences were taken from \cite{bela}.

We then form $\bzbar \to D_{sJ}^+ K^-$ and $D_{sJ}^- \pi^+$ 
candidates and extract the signal using 
the energy difference $\Delta E\,=\,E^{CM}_B-E^{CM}_{\rm beam}$
and beam-constrained mass 
$M_{\rm bc}=\sqrt{(E^{CM}_{\rm beam})^2\,-\,(p^{CM}_B)^2}$;
$E^{CM}_B$ and $p^{CM}_B$ are the energy and momentum
of the $B$ candidate in the center-of-mass (CM) system
and $E^{CM}_{\rm beam}$ is the CM beam energy.
Only events within the intervals 
$M_{\rm bc} > 5.2\,\gev/c^2$ and $|\Delta E|\,<0.2\,\gev$ are used
in this analysis.
The $B$ meson signal region is defined by
$|\Delta E|\,<0.04\,\gev$ and 
$5.272\,\gev/c^2\, < M_{\rm bc} < 5.288\,\gev/c^2$.

Combinatorial background for channels involving the 
$D_{sJ}(2460)^+$ was further suppressed by requiring
\mbox{$\cos \theta_{D_{s}\gamma} < 0.7$}.
The helicity angle $\theta_{D_{s}\gamma}$ is
defined as the angle between 
the direction opposite the $B$ momentum and 
the $D_s^+$ momentum in the $D_s^+ \gamma$ rest frame.
This requirement rejects 49$\%$ of background 
events and only 6$\%$ of signal events, assuming \mbox{$J^P = 1^+$} 
for the $D_{sJ}(2460)^+$. The uncertainty due to
this assumption is included in the systematic error.

For events with two or more $B$ candidates,
the $D_s^+$ and $\pi^0$ candidates
with invariant masses closest to their nominal values
and the $B$ daughter $K^+$ or $\pi^-$ candidate 
with the best ${\cal L}_{K/\pi}$ value are chosen.
With these requirements no multiple entries are
allowed for the $D_{sJ}^\ast(2317)^+$ channels
and, according to the Monte Carlo (MC) simulation,
less than $1\%$ of selected events will have two $B$ candidates
in channels with $D_{sJ}(2460)^+$.
No multiple entries are found in the data.

After this selection the principal background is from 
$e^+e^- \rightarrow q \bar{q}$ continuum events ($q = u,d,s,$ or $c$).
We exploit the event topology to separate $B\bbar$ events (spherical)
from the continuum background (jetlike).
The ratio of the second and zeroth Fox-Wolfram moments~\cite{fox} of all particles
in the event is required to be less than 0.5.
For such events, we form a Fisher discriminant
from six modified Fox-Wolfram moments~\cite{fisher}.
A signal (background) likelihood ${\cal L}_S$ (${\cal L}_{BG}$) is obtained
using signal MC (sideband) data from the product of probability density
functions for the Fisher discriminant and $\cos\theta_B$, where
$\theta_B$ is the $B$ flight direction in the CM system 
with respect to the $z$ axis.
We require  
${\cal R} = {\cal L}_S/({\cal L}_S + {\cal L}_{BG}) > 0.4$ for 
$D_s^+ \to \kzbarstar K^+$ and ${\cal R} > 0.25$ for the other
$D_s^+$ decay modes, which have lower backgrounds.
For the $\bzbar \to D_{sJ}^\ast(2317)^+ K^-$ mode these 
requirements retain 92\%, 85\%, and 95\%
of signal events while removing 
47\%, 67\%, and 64\% of continuum events, 
for $D_s^+ \to \phi \pi^+$, $\kzbarstar K^+$, 
and $K^0_S K^+$, respectively. 
The fractions retained (or removed) for the other $B$ decay modes are similar,
varying by a few percent.


\begin{figure*}[t!]
\vspace{-1.4cm}
\begin{center}
\epsfig{file=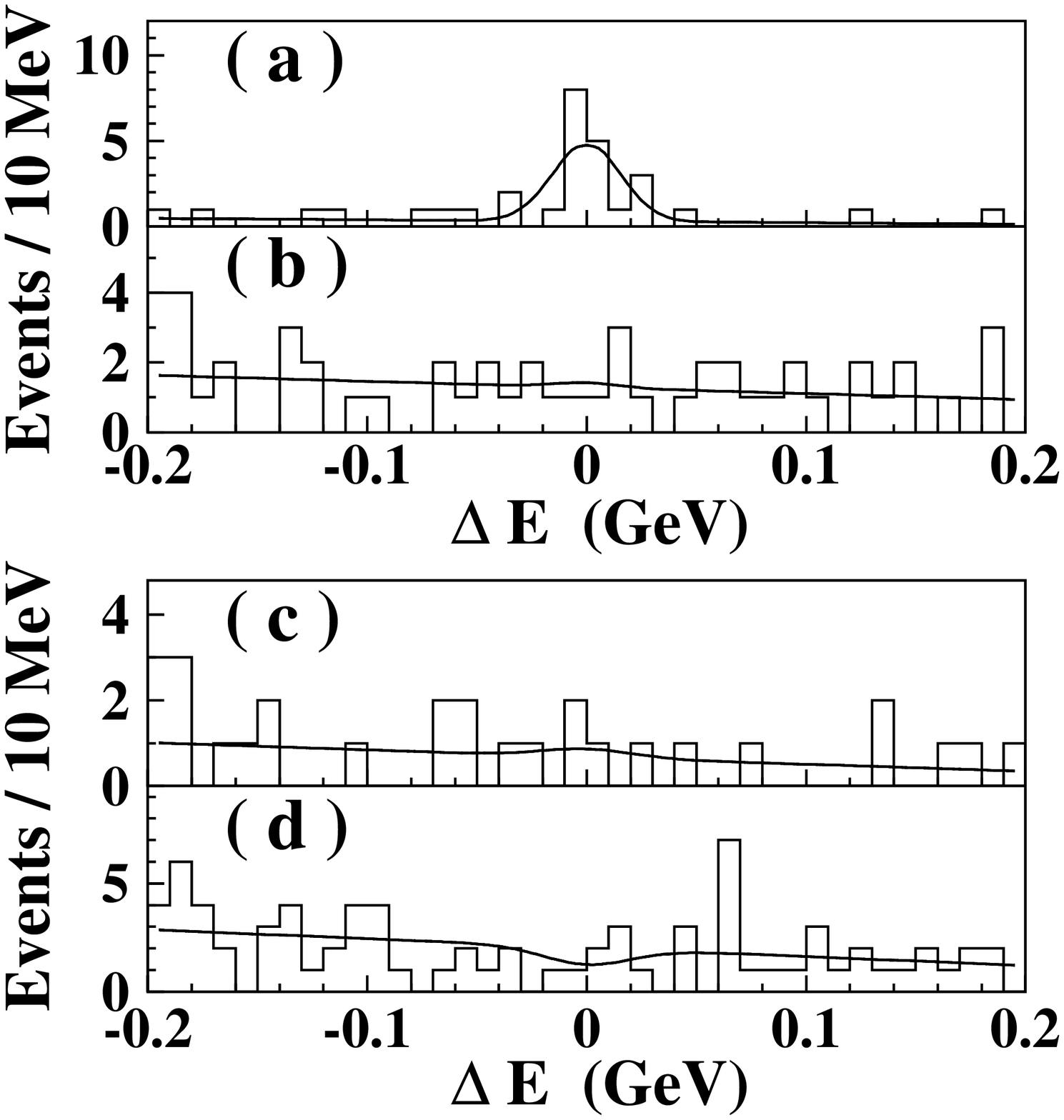,width=8.0cm,height=10.cm}\epsfig{file=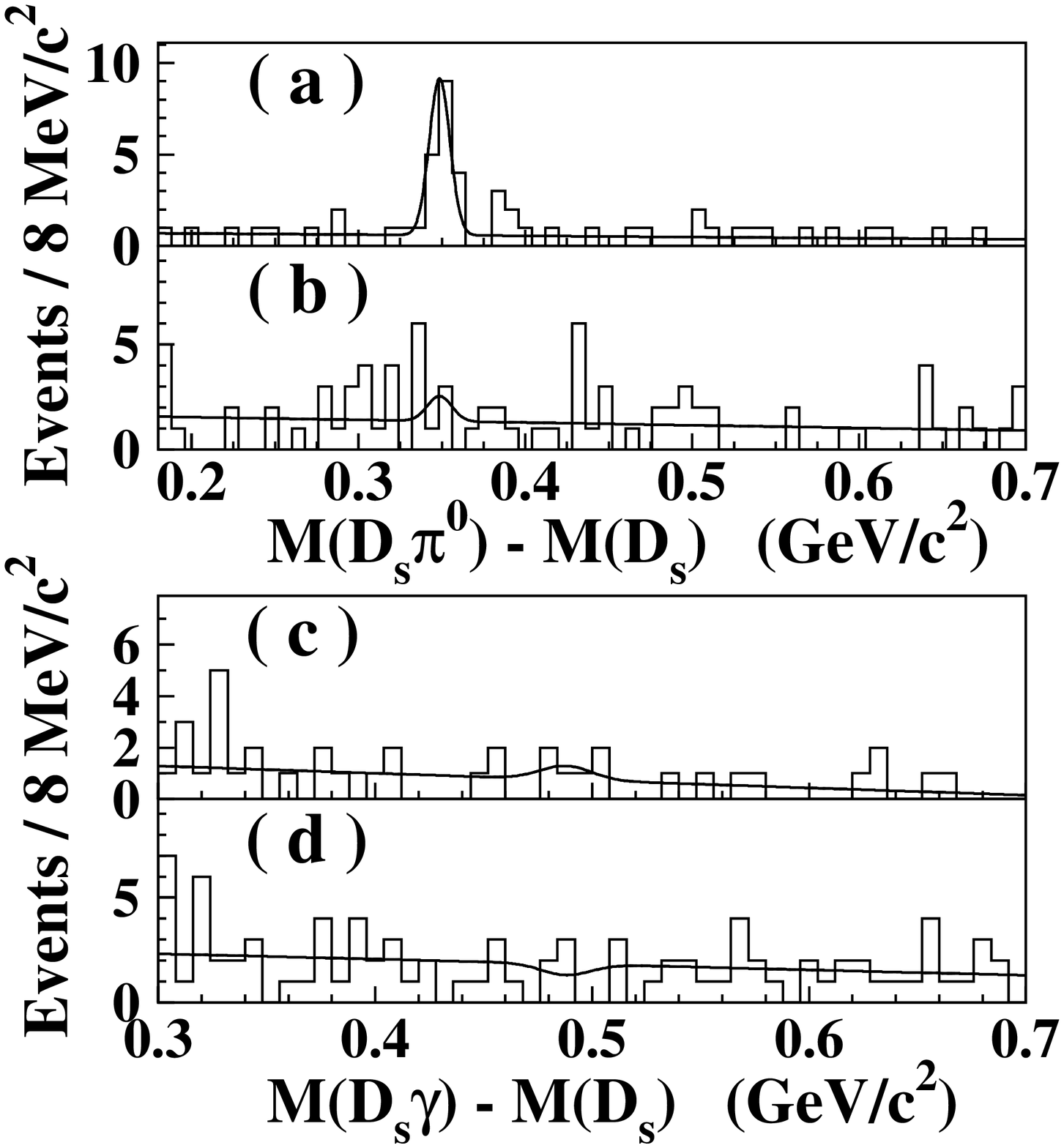,width=8.0cm,height=10.0cm}
\end{center}
\vspace{-0.6cm}
\caption{$\Delta E$ (left) and $\Delta M(D_{sJ})$ (right) distributions
for \bzbar\ decays to 
(a) $D_{sJ}^\ast(2317)^+ K^-$, \mbox{(b) $D_{sJ}^\ast(2317)^- \pi^+$},
(c) $D_{sJ}(2460)^+ K^-$ and (d) $D_{sJ}(2460)^- \pi^+$.
Tight requirements on $M_{\rm bc}$, and $\Delta M(D_{sJ})$ (left) or $\Delta E$ (right) are applied; see the text.}
\label{figall}
\end{figure*}

The $\Delta E$ and $\Delta M(D_{sJ})$ distributions
for the various $D_{sJ}^+ K^-$ and $D_{sJ}^- \pi^+$ combinations
are shown in Fig.~\ref{figall}
for the range $5.272\,\gev/c^2\, < M_{\rm bc} < 5.288\,\gev/c^2$.
To obtain the $\Delta M(D_{sJ})$
distributions we relax the $\Delta M(D_{sJ})$ requirements and
apply a tight selection on $\Delta E$.
Each $\Delta E$ distribution is fitted by a Gaussian 
with zero mean and a width fixed from MC data to describe the signal, and
a linear background function.
The $\Delta M(D_{sJ})$ distributions are described by
signal Gaussians with widths fixed from MC data and mass
differences fixed to $348.6\,\mev/c^2$ or $487.9\,\mev/c^2$, and 
linear backgrounds.
A clear $\bzbar \to D_{sJ}^\ast(2317)^+ K^-$ signal is observed;
no significant signals are observed in the remaining modes
(Fig.~\ref{figall}).
The \bzbar\ yields, based on fits to 
histograms combining all three $D_s^+$ decay modes,
are listed in the last
four lines of Table~\ref{tab:bfr}.

Various studies are performed to confirm the 
$\bzbar \to D_{sJ}^\ast(2317)^+ K^-$ signal.
The parameters of the signal peak 
are allowed to float in the $\Delta M(D_{sJ})$ fit:
a mean $(351.2 \pm 1.6)\,\mev/c^2$ and a width 
$(6.0 \pm 1.2)\,\mev/c^2$ are obtained, 
in good agreement with the MC expectations,
$(348.5 \pm 0.3)\,\mev/c^2$ and $(6.1 \pm 0.2)\,\mev/c^2$.
(The mass of $2317.5\,\mev/c^2$ and the zero width are used in
the MC simulation of the $D_{sJ}^\ast(2317)^+$ signal.)
Good agreement is also obtained for the signal position and width
in $\Delta E$ and $M_{\rm bc}$.

To check for a possible background contribution due to a random 
combination of a $D_{sJ}^\ast(2317)^+$ meson and a kaon,
$\Delta M(D_{sJ}^\ast(2317)^+)$ distributions are obtained for events
from the $\Delta E$
sideband $0.05\,\gev < |\Delta E| <\, 0.2\,\gev$, and the $M_{\rm bc}$
sideband \mbox{$5.2\,\gev/c^2\, < M_{\rm bc} < 5.26\,\gev/c^2$}.
After rescaling the fit results to the $B$ signal region,
the background contribution is estimated to be $-0.8 \pm 0.7$ ($1.1 \pm 0.9$)
events using the $\Delta E$ ($M_{\rm bc}$) sideband.
As both $\Delta E$ and $M_{\rm bc}$ requirements are applied 
in the $\Delta M(D_{sJ}^\ast(2317))$ fit, this background contribution is estimated
to be less than one event, and treated as a source of systematic error.

\renewcommand{\arraystretch}{1.2}
\begin{table*}[t!]
\begin{ruledtabular}
\caption{Signal yields, efficiencies, product branching fractions (or limits),
and significances for the $\bzbar \to D_{sJ}^+ K^-$ and 
$D_{sJ}^- \pi^+$ decay modes. Only statistical errors are shown.
Product branching fractions are obtained from $\Delta M(D_{sJ})$ fits:
see the text.}
\label{tab:bfr}
\begin{tabular}{lBBBdd}
Decay mode	& \multicolumn{1}{c}{Yield} 
                & \multicolumn{1}{c}{Yield}
		& \multicolumn{1}{c}{\phantom{$\pm$}Efficiency}
		& \multicolumn{1}{c}{Product $\mathcal{B}(\bzbar\to D_{sJ}h) \times$}
		& \multicolumn{1}{c}{Signif.}	\\
		& \multicolumn{1}{c}{\phantom{$\pm$}$\Delta M(D_{sJ})$}
		& \multicolumn{1}{c}{\phantom{$\pm$}$\Delta E$}
		& \multicolumn{1}{c}{\phantom{$\pm$}$(10^{-4})$}
		& \multicolumn{1}{c}{$\mathcal{B}(D_{sJ} \to D_s \pi^0 (\gamma))\; (10^{-5})$}
		& \multicolumn{1}{c}{$\sigma$}		 \\ \hline 
$\bzbar\to D_{sJ}^\ast(2317)^+K^-$ &  &	& & & 	\\
\phantom{$\bzbar$} $D_s \to\phi\pi$	& 7.5B^{+3.1}_{-2.5} &	& 8.8 B\pm 0.6	& 5.6^{+2.4}_{-1.9}	& 4.6	\\
\phantom{$\bzbar$}
   $D_s \to \kzbarstar K$ & 3.3B^{+2.6}_{-1.8} & & 7.1 B\pm 0.5	& 3.1^{+2.3}_{-1.7} & 2.3 \\
\phantom{$\bzbar$}
   $D_s \to K_S^0 K$ & 5.7B^{+2.8}_{-2.1} & & 5.8 B\pm 0.5 & 6.6^{+3.2}_{-2.4}	& 4.1	\\
\phantom{$\bzbar$} Simultaneous fit &  &  &  & 5.3^{+1.5}_{-1.3}	& 6.7	\\
\phantom{$\bzbar$} Sum of three modes & 16.6B^{+4.6}_{-4.1} & 17.6B \pm 4.5 & &  & \\ \hline
$\bzbar \to D_{sJ}^\ast(2317)^- \pi^+$		& 2.9B^{+3.3}_{-2.8} & 0.5B \pm3.3 & 27.6 B\pm 1.3	& <2.5\,(90\%\,\text{C.L.})  &	\\ 
$\bzbar \to D_{sJ}(2460)^+ K^-$			& 2.0B^{+2.9}_{-2.2} & 1.0B \pm 2.9 & 56.5 B\pm 2.4	& <0.94\,(90\%\,\text{C.L.}) &	\\ 
$\bzbar \to D_{sJ}(2460)^- \pi^+$		& -1.9B^{+3.1}_{-2.6} & -3.9B \pm 4.1 & 65.6 B\pm 2.6	& <0.40\,(90\%\,\text{C.L.}) &	\\
\end{tabular}
\end{ruledtabular}
\end{table*}

  The shape of the background-subtracted and efficiency-corrected 
$\cos\theta_{D_s \pi}$ distribution for $\bzbar \to D_{sJ}^\ast(2317)^+ K^-$
decay is compared with those predicted 
for possible $D_{sJ}^\ast(2317)^+$ quantum number hypotheses.
(The helicity angle $\theta_{D_{s} \pi}$
is defined as for $\theta_{D_{s} \gamma}$, 
with $\pi^0$ substituted for $\gamma$.)
The 
distribution is expected to be flat if the $D_{sJ}^\ast(2317)^+$ 
has $J^P = 0^+$, or to have the form $\cos^2 \theta_{D_s \pi}$
in the $1^-$ case; 
within large errors, it is consistent with a constant.
A fit gives $\chi^2 = 1.44$ for a constant
and $\chi^2 = 4.72$ for $\cos^2 \theta_{D_s \pi}$,
for four degrees of freedom.
A larger data sample is required for a statistically 
significant separation of the two hypotheses.

Signal yields, efficiencies, branching fractions and significances
for the studied decay channels are shown
in Table~\ref{tab:bfr}.
The $\bzbar \rightarrow D_{sJ}^\ast(2317)^+ K^-$ branching fraction is obtained
using a simultaneous fit to the $\Delta M(D_{sJ}^\ast(2317))$ distributions 
for the three $D_s^+$ decay channels, with independent background descriptions,
but common values for the signal width (fixed from MC)
and peak position (allowed to float).
The branching fraction thus obtained 
is in good agreement with the values from the
$\Delta E$ and $M_{\rm bc}$ fits.
Efficiencies include all intermediate resonance 
branching fractions \cite{pdg} 
and were obtained from MC simulation, 
assuming $J^P = 0^+$ for the $D_{sJ}^\ast(2317)$
and $J^P = 1^+$ for the $D_{sJ}(2460)$.
We assume equal production of neutral and charged $B$ mesons.
The significance is defined as
$\sqrt{-2\,\ln({\cal L}_0/{\cal L}_{\rm max})}$,
where ${\cal L}_{\rm max}$ and ${\cal L}_0$ are likelihoods
(corrected for the number of degrees of freedom)
for the best fit and zero signal yields, respectively.
The upper limits are obtained using fits to
$\Delta M(D_{sJ})$ distributions, with fixed signal positions and widths.
We use the Feldman-Cousins method~\cite{feco},
assuming a Gaussian distribution for the statistical error.
The upper limit is then increased by 29\% (the
sum in quadrature of the experimental systematic
error and the uncertainty in the $D_s^+$ branching fraction scale).
The systematic error is treated in a conservative way in order
to avoid Bayesian assumptions about its probability distributions.
Other methods for upper limit determinations agree
with the values obtained here within 5\% for the first two
upper limits (that have positive signals) and within 15\% for the last
upper limit.

The main result of this study is the measurement 
of the product branching fraction
${\cal B}(\bzbar \rightarrow D_{sJ}^\ast(2317)^+ K^-) \times
{\cal B}(D_{sJ}^\ast(2317)^+ \rightarrow D_s^+ \pi^0) = 
(5.3^{+1.5}_{-1.3} \pm 0.7 \pm 1.4) \cdot 10^{-5}$.
The three error terms are the statistical uncertainty, the total systematic
error, and the uncertainty due to $D_s^+$ branching fractions;
this last term is dominated by the $\sim25\%$ uncertainty
in ${\cal B}(D_s^+ \to \phi \pi^+)$ \cite{pdg}.

The major sources contributing to the systematic error are uncertainties in
efficiencies of charged track
reconstruction (1\% $\times \,N_{\rm tracks}$),
particle identification for charged pions \mbox{(2\% $\times \,N_{\pi}$)}
and kaons \mbox{(2--3\% $\times \,N_{K}$)};
the photon and $\pi^0$ reconstruction efficiencies
and energy scale (5\%);
the $K^0_S$ vertex reconstruction (3\%);
the efficiency of the $\Delta E$ (2\%) and
topological likelihood ratio
($\mathcal{R}$) selections (3\%);
the background and signal shape definition for the $B$ signal (3\%); 
the background subtraction (6\%);
the change in reconstruction efficiency 
for the different $D_{sJ}^+$ quantum number assumptions (4\%);
the statistical uncertainty of the MC sample used to determine 
efficiency (4\%); and the uncertainty 
on the number of $B\bbar$ pairs (0.5\%).
These uncertainties were 
added in quadrature to obtain a total systematic error of 13\%.

In conclusion, the 
$\bzbar \to D_{sJ}^+ K^-$ and $D_{sJ}^- \pi^+$
decay modes were studied for the first time.
The $\bzbar \to D_{sJ}^\ast(2317)^+ K^-$ mode was observed, with a product branching fraction 
\mbox{${\cal B}(\bzbar \rightarrow D_{sJ}^\ast(2317)^+ K^-) \times
{\cal B}(D_{sJ}^\ast(2317)^+ \rightarrow D_s^+ \pi^0) = 
(5.3^{+1.5}_{-1.3} \pm 0.7 \pm 1.4) \cdot 10^{-5}$}.
Recent measurements imply that the
$D_{sJ}^\ast(2317)^+ \to D_s^+ \pi^0$ channel is dominant and the 
$D_{sJ}(2460)^+ \to D_s^+ \gamma$ fraction is around 30$\%$.
Taking into account these approximate values, we can conclude that
$\mathcal{B}(\bzbar \to D_{sJ}^\ast(2317)^+ K^-)$ 
is of the same order of magnitude as $\mathcal{B}(\bzbar \to D_s^+ K^-)$
and at least a factor of two larger than
the $\bzbar \to D_{sJ}(2460)^+ K^-$ branching fraction,
in contrast to the na\"{\i}ve expectation that decays with
the same spin-doublet 
$D_{sJ}^\ast(2317)^+$ and $D_{sJ}(2460)^+$ mesons would have similar rates.
No significant signals for $\bzbar \to D_{sJ}^- \pi^+$ decays were seen.

We thank the KEKB group for the excellent operation of the
accelerator, the KEK cryogenics group for the efficient
operation of the solenoid, and the KEK computer group and
the NII for valuable computing and Super-SINET network
support.  We acknowledge support from MEXT and JSPS (Japan);
ARC and DEST (Australia); NSFC (contract No.~10175071,
China); DST (India); the BK21 program of MOEHRD and the CHEP
SRC program of KOSEF (Korea); KBN (contract No.~2P03B 01324,
Poland); MIST (Russia); MESS (Slovenia); Swiss NSF; NSC and MOE
(Taiwan); and DOE (USA).


\begin{thebibliography}{99}
\bibitem{fna}
Charge conjugate modes are implicitly included everywhere.
\bibitem{fnb}
In the following description all momenta are stated 
in the laboratory frame unless otherwise specified.
\bibitem{zdef}
The $z$ axis is antiparallel to the positron beam direction.
\end{thebibliography}
\end{document}